\documentclass[twocolumn]
{IEEEtran} 

\usepackage{amsmath,amsfonts,amsthm}
\usepackage[final]{graphicx}
\usepackage{xcolor}

\usepackage{subfig}

\usepackage{amssymb,amsfonts}

\usepackage{balance}
\usepackage{fancyhdr}

\usepackage{bookmark}
\usepackage{verbatim}
\usepackage{fontenc}
\usepackage{cuted}
\usepackage{graphicx} 
\usepackage{float}
\usepackage{array} 
\usepackage{paralist} 
\usepackage[numbers,sort&compress]{natbib}
\usepackage{bm}

\usepackage{fancyhdr}												
\renewcommand{ }{ \textbf}

\newcommand{\ket}[1]{\ensuremath{\left|#1\right\rangle}} 
\newcommand{\bra}[1]{\ensuremath{\left\langle#1\right|}} 

\begin{document}
\title{A Quantum Implementation Model for Artificial Neural Networks} 
 \author{Ammar~Daskin~
 \thanks{A. Daskin is with the Computer Engineering Department, Istanbul Medeniyet University, Istanbul, Turkey, email: adaskin25@gmail.com}}
\maketitle
\begin{abstract}
The learning process for multi layered neural networks  with many nodes makes heavy demands on computational resources. 
In some neural network models, the learning formulas, such as the Widrow-Hoff formula, do not change the eigenvectors of the weight matrix while flatting the eigenvalues. 
In infinity, this iterative formulas result in terms formed by the principal components of the weight matrix: i.e., the eigenvectors corresponding to the non-zero eigenvalues. 

In quantum computing, the phase estimation algorithm is known to provide speed-ups over the conventional algorithms for the eigenvalue-related problems. 
Combining the quantum amplitude amplification with the phase estimation algorithm, a quantum implementation model for artificial neural networks using the Widrow-Hoff learning rule is presented. 
The complexity of the model is found to be linear in the size of the weight matrix. This provides a quadratic improvement over the classical algorithms.
\end{abstract}

\section{Introduction and Background}
Artificial neural networks (ANN) \cite{Haykin2009neural,Abdi2001linear,Abdi1995more} are adaptive statistical models which mimic the neural structure of the human brain to find optimal solutions for multivariate problems.
In the design of ANN, the followings are determined: the structure of the network, input-output variables, local activation rules, and a learning algorithm.
 Learning algorithms are generally linked to the activities of neurons and describe a mathematical cost function.
Often, a minimization of this cost function composed of the weights and biases describes the learning process in artificial neural networks. 
Moreover,  the learning rule in this process specifies how  the synaptic weights should be updated at each iteration. 
In general, learning rules can be categorized as supervised and unsupervised learning: In supervised learning rules, 
the distance between the response of the neuron and a specified response, called target $t$, is considered. However, it is not required in unsupervised learning rules. 
Hebbian learning rule\cite{Morris1999437} is a typical example of the unsupervised learning,
in which the weight vector at the $(j+1)$th iteration is updated by the following formula (We will mainly follow Ref.\cite{Abdi2001linear} to describe learning rules.): 
\begin{equation}
\mathbf{w_{[j+1]}}  = \mathbf{w_{[j]}} -\eta t\mathbf{x}. 
\end{equation}
Here, $\mathbf{x}$ is the input vector, $\eta$ is a positive learning constant and $\mathbf{w_{[j]}}$ represents the weights at the $j$th iteration. \textbf{And $t$ is the target response. Learning is defined by getting an output closer to the target response.}

On the other hand,  Widrow-Hoff learning rule\cite{Widrow1960adaptive}, which is the main interest of this paper, illustrates a typical supervised learning rule \cite{Abdi1995more,Abdi1996widrow,Abdi2001linear}:
\begin{equation}
\mathbf{w_{[j+1]}} = \mathbf{w_{[j]}} -\eta \sigma'(v) (t-y) \mathbf{x}, 
\end{equation}
where $v=\mathbf{x}^T\mathbf{w}$ is the activation of the output cell and $\sigma'(v)$ is the derivative of the activation function which specifies the output of a cell in the considered network, $y = \sigma(v)$: e.g., the sigmoid function, $\sigma(v)=1/(1+exp(-v))$.
\textbf{While in the Hebbian iteration the weight vector is moved in the direction
	of the input vector by an amount proportional to the target, in the Widrow-Hoff iteration, the change is proportional to the error $(t-y)$.}
If we consider multi-neurons; the activation, the output, and the target values becomes vectors: viz.,  $\mathbf{v, y}$ and $\mathbf{t}$, respectively. 
\textbf{When there are several input and target associations, the set of inputs, targets, activations, and outputs can be represented by the matrices $X,T,V,$ and $Y$, respectively. Then, the above equations come in matrix forms as follows:} 
\begin{equation}
W_{[j+1]} = {W_{[j]}} -\eta XT^T, 
\end{equation}
\begin{equation}
\label{EqWidrowHoffVec}
W_{[j+1]} = W_{[j]} -\eta (\sigma'(V) \circledast X )(T-Y)^T, 
\end{equation}
where $W$ represents the matrix of synaptic weights.

It is known that the learning  task for multi layered neural networks  with many nodes  makes heavy demands on computational resources. 
Algorithms in quantum computational model provide computational speed-up over their classical counterparts for some particular problems: e.g., Shor's factoring algorithm\cite{Shor1994algorithms} and Grover's search algorithm\cite{Grover1996fast}.  
Using adiabatic quantum computation\cite{Neven2008training,Neven2009training} or mapping data set to  quantum random access memory \cite{Aimeur2013quantum, Lloyd2014topological} speed-ups in big data analysis have been shown to be possible  \cite{Rebentrost2014quantum, Wittek2014quantum, Schuld2014quantum}.  Furthermore, Lloyd et al.\cite{lloyd2014qQPCA} have described a quantum version for principal component analysis.

In the recent decades, particularly relating the neurons in the networks with qubits \cite{Manju2014}, a few different quantum analogous of the artificial neural networks have also been developed: 
e.g.\cite{DaSilva201655,Zhou2012,GUPTA2001355,
Andrecut2002quantum,Altaisky2001quantum,Schuld2014QW} (For a complete review and list of references, please refer to Ref.\cite{Schuld2014quest}).  
These models should not be confused with the classical algorithms (e.g. see Ref.\cite{Kouda2005qubit,Li2013hybrid}) inspired by the quantum computing.  
Furthermore, using the Grover search algorithm \cite{Grover1996fast}, a quantum associative memory is introduced \cite{Ventura1999}.
Despite some promising results, there is still need for further research on new models\cite{Schuld2014quest}.

The quantum phase estimation algorithm (PEA)\cite{Kitaev}  provides computational speed-ups over the known classical algorithms in eigenvalue related problems. 
The algorithm mainly finds the phase value of the eigenvalue of a unitary matrix (considered as the time evolution operator of a quantum Hamiltonian)  for a given approximate eigenvector.
Because of this property, PEA is ubiquitously used as a subcomponent of other algorithms. 
While in the general case, 
PEA requires a good initial estimate of an eigenvector to produce the phase; in some cases, it is able to find the phase by using an  initial equal superposition state: e.g., Shor's factoring algorithm \cite{Shor1994algorithms}. 
In Ref.\cite{Harrow2009quantum},  it is shown that  a flag register can be used in the phase estimation algorithm to eliminate the ill-conditioned part of a matrix by processing the eigenvalues greater than some threshold value.
Amplitude amplification algorithm 
\cite{Grover1996fast,Grover1998,Mosca1998quantum, Brassard2002} is used to amplify amplitudes of certain chosen quantum states considered.
In the definition of  quantum reinforcement learning \cite{Chen2006quantum,Dong2008QRL}, states and actions are represented as quantum states. And based on the observation of states a reward is applied to the register representing actions. Later, the quantum amplitude amplification is applied to amplify the amplitudes of rewarded states. 
In addition, in a prior work \cite{Daskin2016QPCA} combining the amplitude amplification with the phase estimation algorithm, we have showed a framework to obtain the eigenvalues in a given interval and their corresponding eigenvectors from  an initial equal superposition state.  
This framework can be used as a way of doing quantum principal component analysis (QPCA).

For a given weight matrix $W$;  in \textbf{linear auto-associators} using the Widrow-Hoff learning rule; during the learning process, the eigenvectors does not change while the eigenvalues goes to one \cite{Abdi1996widrow,Abdi2001linear}: i.e.,
$lim_{j\rightarrow\infty} W_{[j]}$ 
converges to $QQ^T$, where $Q$ represents the eigenvectors of $W$. 
Therefore, for a given input $\mathbf{x}$, the considered network produces the output $QQ^T\mathbf{x}$.  
In this paper, we present a quantum implementation model for the artificial neural networks 
by employing the algorithm in Ref.\cite{Daskin2016QPCA}. 
In particular, we show how to construct $QQ^T\mathbf{x}$ on quantum computers in linear time. In the following section, we give the necessary description of Widrow-Hoff learning rule and QPCA described in Ref.\cite{Daskin2016QPCA}. 
In Sec.\ref{section3},  we shall show how to apply QPCA to the neural networks given by the Widrow-Hoff learning rule and discuss  the possible implementation issues such as the circuit implementation of $W$, the preparation of the input $\mathbf{x}$ as a quantum circuit, and determining the number of iterations in the algorithm. In Sec.\ref{section4}, we analyze the complexity of the whole application. Finally, in Sec.\ref{SecExample}, an illustrative example is presented.

\section{Methods}
\label{section2}
In this section, we shall describe the Widrow-Hoff learning rule and the quantum algorithms used in the paper.
\subsection{Widrow-Hoff Learning}
\textbf{For a linear autoassociator, i.e. $Y=V$, $\sigma'(V) = I$, and $T=X$;} Widrow-Hoff learning  rule given in Eq.\eqref{EqWidrowHoffVec}, also known as LMS algorithm, 
in matrix form can be described 
as follows \cite{Abdi1995more,
Abdi2001linear}:
\begin{equation}
W_{[j]}=W_{[j-1]}+\eta (X-W_{[j-1]}X)X^T.
\end{equation}
This can be also expressed by using the eigendecomposition of $W=Q\Lambda Q^T$: i.e., $W_{[j]}=Q\Phi_{[j]} Q^T$, where $\Phi_{[j]}=[I-(I-\eta\Lambda)^j]$. 
$\Phi_{[j]}$ is called the eigenvalue matrix at the epoch $j$. Based on this formulation,  Widrow-Hoff error correction rule only affects the eigenvalues  and flattens them when $\eta \leq 2\lambda_{max}^{-1}$ ($\lambda_{max}$ is the largest eigenvalue of $W$): i.e., $\lim_{j\rightarrow\infty}\Phi_{[j]}=I$. Thus, in infinity, the learning process $W$ ends up as: $W_{[\infty]}=QQ^T$.

\subsection{Quantum Algorithms Used in the Model}
   In the following, we shall first explain two well-known quantum algorithms and then describe how they are used in Ref.\cite{Daskin2016QPCA} to obtain the linear combination of the eigenvectors.
   
\subsubsection{Quantum Phase Estimation Algorithm}
\begin{figure}
\centering
\includegraphics[width=3.5in]{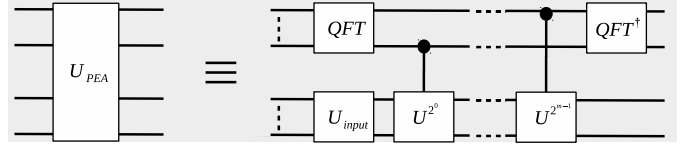}
\caption{The phase estimation part of the algorithm.}
\label{FigPEA}
\end{figure}

The phase estimation algorithm (PEA)
\cite{Kitaev,Nielsen2010quantum} finds an estimation for the phase of an eigenvalue of a given operator.    
In mathematical terms, the algorithm  seen in Fig.\ref{FigPEA} as a circuit works as follows:
\begin{itemize}
\item  An estimated eigenvector \ket{\varphi_j} associated to the $j$th eigenvalue $e^{i\phi_j}$ of a unitary matrix, $U$ of order $N$ is assumed given. $U$ is considered as a time evolution operator of the  Hamiltonian ($H$) representing the dynamics of the quantum system:
\begin{equation}
\label{EqHamiltonian}
U=e^{itH/\hbar},
\end{equation} 
where $t$ represents the time and $\hbar$ is the Planck constant. 
As a result, the eigenvalues of $U$ and $H$ are related: while $e^{i\phi_j}$ is the eigenvalue of $U$, its phase $\phi_j$ is the eigenvalue of $H$.

\item The algorithm uses two quantum registers dedicated to the eigenvalue and the eigenvector, respectively, \ket{reg_1} and \ket{reg_2} with $m$ and $(n=log_2N)$ number of qubits. 
The initial state of the system is set to 
\ket{reg_1}\ket{reg_2}=\ket{\mathbf 0}\ket{\varphi_j}, 
where \ket{\mathbf 0} is the first standard basis vector.

\item Then, the quantum Fourier transform
is applied to \ket{reg_1}, which produces the following equal superposition state: 
\begin{equation}
U_{QFT}\ket{reg_1}\ket{reg_2} = \frac{1}{\sqrt{M}}\sum_{k=0}^{M-1}\ket{\mathbf k}\ket{\varphi_j},
\end{equation}   
where $M=2^m$ and \ket{\mathbf k} is the $k$th standard basis vector.
 \item For  each $k$th qubit in the first register, a quantum operator, $U^{2^{k-1}}$, controlled by this qubit is applied to the second register. This operation leads the first register to hold the discrete Fourier transform of the phase, $\phi_j$.
 \item  The inverse quantum Fourier transform on  the first register produces the binary digits of $\phi_j$.
 \item Finally, the phase is obtained by measuring the first register.
\end{itemize}

\subsubsection{Quantum Amplitude Amplification Algorithm}

If a given quantum state \ket{\psi} in N-dimensional Hilbert space can be rewritten in terms of some orthonormal states considered as the ``good" and the ``bad" parts of \ket{\psi}
as: 
\begin{equation} 
 \ket{\psi} = sin(\theta) \ket{\psi_{good}} +cos(\theta) \ket{\psi_{bad}},
\end{equation} 
then amplitude amplification technique  
\cite{Grover1996fast,Brassard1997exact,Brassard2002quantum} can be used to increase the amplitude of \ket{\psi_{good}} in magnitude while decreasing the amplitude of \ket{\psi_{bad}}. 
The technique mainly consists of two parts: the marking and the amplifying implemented by two operators, respectively $U_f$ and $U_{\psi}$. 
Here, $U_f$ marks-flips the sign of-the amplitudes of \ket{\psi_{good}}  and does nothing to \ket{\psi_{bad}}. $U_f$ can be implemented as a reflection operator when  \ket{\psi_{good}} and \ket{\psi_{bad}} are known:
\begin{equation}
U_f = I- 2\ket{\psi_{good}}\bra{\psi_{good}},
\end{equation}
where $I$ is an identity matrix.
In the amplification part, the marked amplitudes are amplified by the application of the operator $U_{\psi}$:
\begin{equation}
U_{\psi} = I-2\ket{\psi}\bra{\psi}
\end{equation}
To maximize the probability of \ket{\psi_{good}}, 
the iteration operator $G= U_{\psi}U_f$ is iteratively $O(\sqrt{N})$ times applied to the resulting state.

\subsection{Quantum principal component analysis}

In Ref.\cite{Daskin2016QPCA}, we have shown that combining PEA with the amplitude amplification, one can obtain eigenvalues in  certain intervals. 

In the phase estimation part, 
the initial state of the registers is set to
\ket{\mathbf 0}
\ket{\mathbf 0}. 
Then, the second register is put into 
the equal superposition state $1/\sqrt{N}(1, \dots, 1)^T$. 
The phase estimation process in this input generates the superposition of the eigenvalues on the first  and the eigenvectors on the second register. 
In this final superposition state, 
the amplitudes for the eigenpairs  are proportional to the norm of the projection of the input vector onto the eigenvector: i.e., the normalized sum of the eigenvector elements.
This part is represented by $U_{PEA}$ and also involves the input preparation circuit, $U_{input}$, on the second register. 

In the amplification part, first,  $U_f$ is applied to the first register to mark the eigenvalues determined by the binary values of the eigenvalues: For instance, if we want to mark an eigenvalue equal to 0.25 in \ket{reg_1} with 3 qubits, 
we use $U_f= I-2\ket{010}\bra{010}$ since the binary form of 0.25 is (010) \textbf{(the left most bit represents the most significant bit)}.
The amplitudes of the marked eigenvalues are then amplified by  the application of $U_{\psi}$ with $\ket{\psi}$ representing the output of the phase estimation: 
 \begin{equation}
\ket{\psi} = U_{PEA}\ket{reg_1}
\ket{reg_2}= U_{PEA}\ket{\mathbf 0}
\ket{\mathbf 0}.
\end{equation}
Using the above equation, 
$U_{\psi}$ can be implemented as:  
\begin{equation}
\label{EqUpsi}
U_{\psi}=I-2\ket{\psi}\bra{\psi} = U_{PEA}U_0U_{PEA}^\dagger,
\end{equation}
where $U_0=I-2\ket{\bm 0}\bra{\bm 0}$. 
The amplitudes of the eigenvalues in the desired region are further amplified by the iterative application of the operator $G=U_{\psi}U_f$.
At the end of this process,  a linear combination of the eigenvectors with the coefficients determined by the normalized sum of the vector elements of the eigenvectors are produced.
In the following section, we shall show how to apply this process to model the implementation of the neural networks based on the  Widrow-Hoff learning rule. 

\section{Application to the Neural Networks}
\label{section3}
Since the weight matrix in Widrow-Hoff learning rule converges to the principal components in infinity\cite{Abdi1996widrow}: i.e., $W_{[\infty]}=QQ^T$, the behavior of the trained network on some input \ket{\mathbf x} can be concluded as:
\begin{equation}
W_{[\infty]}\ket{\mathbf x}=QQ^T\ket{\mathbf x}.
\end{equation}
Our main purpose is to find an efficient way to implement this behavior on quantum computers by using the quantum principal component analysis.
For this purpose, we form  $U_f$ in a way that marks 
only the non-zero eigenvalues and their corresponding eigenvectors:
For zero eigenvalues ( in binary form $(0\dots0)$ ), the first register is in $\ket{\mathbf 0} = (1,0,0\dots,0,0)^T$ state. Therefore, we need to construct a $U_f$ which ``marks" the nonzero eigenvalues and does nothing to $\ket{\mathbf 0}$. This can be done by using a vector  \ket{\mathbf f} in the standard basis which has the same non-zero coefficients for the all basis states except the first one:
\begin{equation}
U_f= I-2\ket{\mathbf f}\bra{\mathbf f}, 
\text{ with } \ket{\mathbf f} = \frac{1}{\mu} \left(\begin{matrix}0\\ 1\\ 1\\ \vdots\\ 1\end{matrix}\right).
\end{equation}
Here, $\mu$ is a normalization constant equal to $\frac{1}{\sqrt{M-1}}$. $U_f$ does nothing when the first register in  \ket{\mathbf 0} state; however, it does not only flip the signs  but also changes the amplitudes of the other states. 
Then, $U_{\psi}$ is applied for the amplification of the marked amplitudes. The iterative application of $U_{\psi}U_f$ results a quantum state where the amplitude of \ket{\mathbf 0} becomes almost zero and the amplitudes of the other states becomes almost equal. At this point, the second register holds $QQ^T\ket{\mathbf x}$ which is the expected output from the neural network.
This is explained in more mathematical terms below.

 \subsection{Details of the Algorithm}
Here, we assume that $U=e^{iWt}$ is given: Later, in Sec.\ref{SecCircuitW}, we shall also discuss how $U$ may be obtained as a quantum circuit from a given $W$ matrix. 
 
\begin{figure}
\centering
\includegraphics[width=3.5in]{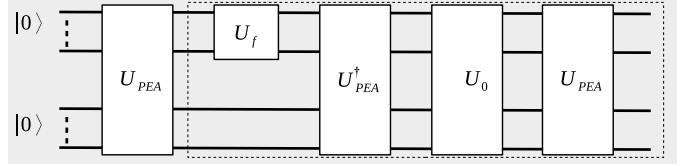}
\caption{The general quantum algorithm to find the principal components of $W$.The dashed box indicates an iteration of the amplitude amplification.}
\label{FigQANN}
\end{figure}

Fig.\ref{FigQANN} shows the algorithm as a quantum circuit where the dashed lines indicates an iteration in the amplitude amplification. 
At the beginning, $U_{PEA}$ is applied to the initial state  \ket{\mathbf 0}\ket{\mathbf 0}. 
Note that
$U_{PEA}$ includes also an input preparation circuit, $U_{input}$, bringing the second register from \ket{\mathbf{0}} state to the input \ket{\mathbf{x}}.
$U_{PEA}$ generates a superposition of the eigenvalues and associated eigenvectors, respectively,  on the first and the second registers with the amplitudes defined by  the overlap of the eigenvector and the input $\ket{\mathbf{x}}$:
\begin{equation}
\ket\psi = U_{PEA}\ket{\mathbf 0}\ket{\mathbf 0}=\sum_{j=0}^{N-1}\alpha_j \ket{\lambda_j}\ket{\varphi_j},
\end{equation} 
 where $\alpha_j =\bra{\varphi_j}\ket{\mathbf x}$. 

\begin{figure}
\centering
\includegraphics[width=3.5in]{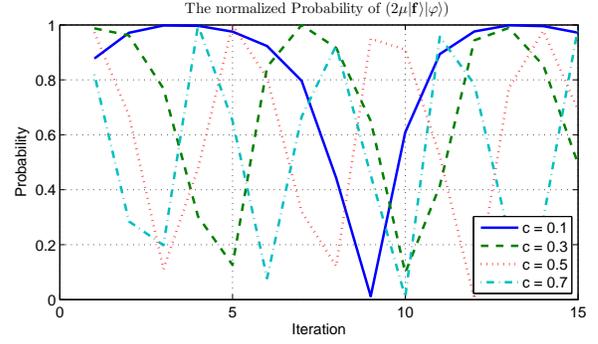}
\caption{The normalized probability of $(2\mu\ket{\mathbf{f}}\ket{\varphi})$ through the iterations for different values of $c$.}
\label{Figmufphi}
\end{figure}

In the second part, the operator $G=U_{\psi}U_f$ is applied to \ket{\psi} iteratively until  $QQ^T\ket{\mathbf{x}}$ can be obtained on the second register.
The action of $U_f$ applied to \ket{\psi} is as follows:
\begin{equation}
\label{EqUfpsi}
\begin{split}
\ket{\psi_1}& =U_f \ket{\psi} 
= \left(I-2\ket{\mathbf f}\bra{\mathbf f}\right) \sum_{j=0}^{N-1}\alpha_j \ket{\lambda_j}\ket{\varphi_j}\\
& =\ket{\psi} -2\mu \ket{\mathbf{f}}\ket{\bar{\varphi}}.
\end{split}
\end{equation}
Here, assuming the first $k$ number of eigenvalues are zero,  
the \underline{unnormalized} state \ket{\bar{\varphi}} is defined as:
\begin{equation}
\ket{\bar{\varphi}} = \sum_{j=k}^{N-1}\alpha_j \ket{\varphi_j}.
\end{equation}
It is easy to see that $\ket{\bar{\varphi}}=QQ^T\ket{\mathbf x}$, which is our target output. 
When $U_{\psi}$ is applied to the output in Eq.\eqref{EqUfpsi}, we simply change the amplitudes of $\ket{\psi}$:
\begin{equation}
\begin{split}
U_{\psi}\ket{\psi_1} & = U_{\psi} \left(\ket{\psi} -2\mu \ket{\mathbf{f}}\ket{\bar{\varphi}}\right)\\
& = \left(I-2\ket{\psi}\bra{\psi}\right)\ket{\psi} -2\mu \left(I-2\ket{\psi}\bra{\psi}\right)\ket{\mathbf{f}}\ket{\bar{\varphi}}\\
&= -\ket{\psi} -2\mu \left(I-2\ket{\psi}\sum_{j=0}^{N-1}\alpha_j \bra{\varphi_j}\bra{\lambda_j}\right)\ket{\mathbf{f}}\ket{\bar{\varphi}}\\
&=-\ket{\psi} -2\mu\ket{\mathbf{f}}\ket{\bar{\varphi}} + 4\mu^2P_f \ket{\psi}\\
& = (4\mu^2P_f -1) \ket{\psi} -2\mu\ket{\mathbf{f}}\ket{\bar{\varphi}}.
\end{split}
\end{equation}
Here, $P_f$ is the initial success probability and equal to 
$\sum_{j=k}^{N-1}\alpha_j^2$.
The repetitive applications of $G$ only changes the amplitudes of $\ket{\psi}$ and $\ket{\mathbf{f}}\ket{\bar{\varphi}}$: e.g., 
\begin{equation}
\begin{split}
G^2\ket{\psi} = (c^2-3c+1)\ket{\psi} -(c-2) 2\mu\ket{\mathbf{f}}\ket{\bar{\varphi}}\\
G^3\ket{\psi} = (c^3-5c^2+ 6c-1)\ket{\psi} -(c^2-4c+3) 2\mu\ket{\mathbf{f}}\ket{\bar{\varphi}}
\end{split}
\end{equation} 
where $c=(4\mu^2P_f -1)$.   The normalized probability of $(2\mu\ket{\mathbf{f}}\ket{\varphi})$ is  presented in Fig. \ref{Figmufphi} by using different values for $c$ (The amplitudes of \ket{\psi} and ($2\mu\ket{\mathbf{f}}$) are normalized.).  The amplitude of \ket{\psi} through the iterations of the amplitude amplification oscillates  with a frequency depending on the overlaps of the input with the eigenvectors. 
When the amplitude of $\ket{\psi}$ becomes close to zero, the second register in the remaining part $\ket{\mathbf{f}}\ket{\bar{\varphi}}$ is exactly $QQ^T\ket{\mathbf x}$ and the first register is equal to $\ket{\mathbf{f}}$. 

Fig.\ref{FigRandomQANN} represents the iterations of the algorithm for a random $2^7 \times 2^7$  matrix with $2^7/2$ number of zero eigenvalues and a random input $\ket{\mathbf x}$ (MATLAB code for the random generation is given in Appendix \ref{AppRandom}). In each subfigure, we have used different numbers of qubits for the first register to see the effect on the results. 
The bar graphs in the subfigures shows the probability change for each state \ket{\mathbf j}, $j=0\dots 1$, of the first qubit (A different color tone indicates a different state.). 
When the probability for \ket{\mathbf 0} becomes close to zero, the probability for the rest of the states becomes equal 
and so  the total of these probabilities as shown in the bottom figure of each subfigure becomes almost one. 
 At that point, the fidelity found by $\left|\bra{reg_2}QQ^T\ket{\mathbf x}\right|$ also comes closer to one.

\subsection{Number of Iterations}
\begin{figure}
\centering
\includegraphics[width=3.5in]{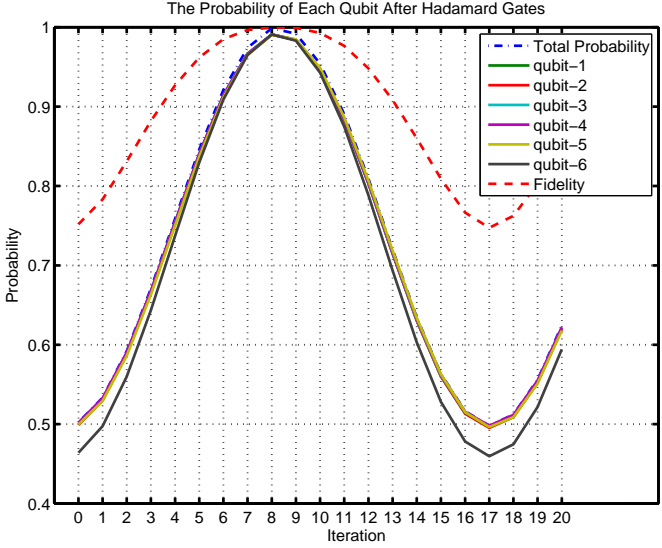}
\caption{The probability to see a qubit in the first register in \ket{0} state after applying a Hadamard gate to the qubit and its comparison with the total probability and the fidelity given in Fig.\ref{FigRandomQANNm6}. Note that the above separate curve is the fidelity. Since there are only small differences between the probabilities on the individual qubits and the total probability, the curves for the probabilities mostly overlap.}
\label{FigRandomANNNqubit}
\end{figure}

Through the iterations, 
while the  probability for \ket{\mathbf 0} state  goes to zero, the probabilities for the rest of the states become almost equal. 
This indicates that the individual states of each  qubit  turns into the equal superposition state. 
Therefore,
if the state of a qubit in the first register is in the  almost equal superposition state, 
then the success probability is very likely to be in its maximum level. 
 In the Hadamard basis, \ket{0} and $\ket{1}$ are represented in the equal superposition states as follows:
\begin{equation}
   \ket{0}= \frac{\ket{0} + \ket{1}}{\sqrt{2}} \text{ and }
   \ket{1}= \frac{\ket{0} - \ket{1}}{\sqrt{2}}. 
\end{equation}  
Therefore, using the Hadamard basis, if the probability of measuring $\ket{0}$ is close to one, in other words, if \ket{1} is not seen in the measurement, then the second register likely holds $QQ^T\ket{\mathbf x}$ with a maximum possible fidelity. 
Fig.\ref{FigRandomANNNqubit} shows the comparisons of the individual qubit probabilities (i.e., the probability to see a qubit in the first register in \ket{0} in the Hadamard basis.) with the total probability observed in Fig.\ref{FigRandomQANNm6} for the random case:
As seen in the figure, the individual probabilities exhibit the same behavior as the total probability.

Generally, obtaining a possible probability density  of an unknown quantum state is a difficult task. 
However, since we are dealing with only a single qubit and does not require the exact density, this can be done efficiently. For instance, if \ket{0} is seen $a$ number times in ten measurements, then the success probability is expected to be $a$/$10$. Here, the number of measurements obviously determines the precision in the obtained probability which may also affect the fidelity.  
 
\subsection{Error-Precision (Number of Qubits in \ket{reg_1})}
The number of qubits, $m$, in the first register  should be sufficient to distinguish very small nonzero eigenvalues from the ones which are zero. 
In our numerical random experiments, we have observed that choosing only six or five qubits are enough to get very high fidelity while not requiring a high number of iterations.
The impact of the number of qubits on the fidelity and the probability  is shown in Fig.\ref{FigRandomQANN} in which each sub-figure is drawn by using different register sizes for the same random case.
As seen in the figure, the number of qubits also affects the required number of iterations: e.g., while for $m=3$, the highest fidelity and probability  are seen at the fourth iteration; for $m=6$, it happens  around the ninth iteration. 

\subsection{Circuit Implementation of $W$}
\label{SecCircuitW}
The circuit implementation of $W$ requires forming a quantum circuit representing the time evolution of $W$: i.e., $U=e^{i2\pi Wt}$.  When $W$ is a sparse matrix, the circuit can be formed  by following the method in Ref.\cite{Berry2007sparse}.
However, when it is not sparse but in the following form $W = \sum_j \mathbf{x_{j}}\mathbf{x_{j}}^T$, then the exponential becomes equal to:
\begin{equation}
\label{EqExpW}
U=e^{i2\pi Wt}=e^{i2\pi t\sum_j \mathbf{x_{j}}\mathbf{x_{j}}^T}.
\end{equation}
To approximate the above exponential,
we apply the Trotter Suzuki formula \cite{Trotter1959product,Suzuki1976generalized,Hatano2005finding,Poulin2015} to decompose Eq.\eqref{EqExpW} into the terms $U_j = e^{i2\pi t\mathbf{x_{j}}\mathbf{x_{j}}^T}=U_{\mathbf{x_j}} \bar{I} U_{\mathbf{x_j}}^\dagger$, 
where $\bar{I}$ is a kind of identity matrix with the first element set to $e^{i2\pi t}$, 
and
$U_{\mathbf{x_j}}$ is a unitary matrix with the first row and column equal to $\mathbf{x_j}$. 
For instance, if  the second order Trotter-Suzuki decomposition is applied to Eq.\eqref{EqExpW} (Note that the order of the decomposition impacts the accuracy of the approximation.), the following is obtained: 
\begin{equation}
e^{i2\pi t\sum_{j=1}^\kappa \mathbf{x_{j}}\mathbf{x_{j}}^T} \approx U_j 
\left(
e^{i2\pi \frac{t}{2}\sum_{j=2}^\kappa \mathbf{x_{j}}
\mathbf{x_{j}}^T}
\right)
 U_j.
\end{equation}
Then, the same decomposition is applied to the term $e^{i2\pi t/2\sum_{j=2}^\kappa \mathbf{x_{j}}\mathbf{x_{j}}^T}$ in the above equation. 
This recursive decomposition yields an approximation composed of $(4\kappa)$ number of $U_{\mathbf{x_j}}$ matrices.  Any $U_{\mathbf{x_j}}$ can be implemented as a Householder matrix by using $O(2^n)$ quantum operations  which is linear in the size of $\mathbf{x_j}$ 
\cite{Ivanov2006engineering,Urias2015householder,Ivanov2008synthesis,Bullock2005asym}. 

\subsection{Obtaining a solution from the output}
Generally, the amplitudes of the output vector (the final state of the second register) encodes the information needed for the solution of the considered problem. 
Since obtaining the full density of a quantum state is  known
to be very inefficient for larger systems, one needs to drive efficient measurement schemes specific to the problem. 
For instance, for some problems, comparisons of the peak values instead of the whole vectors may be enough to gauge a conclusion:  
In this case, since a possible outcome in a measurement would be the one with an amplitude likely to be greater than most of the states in magnitude, 
the peak values can be obtained efficiently. 
However, this alone may not be enough for some applications.

\textbf{Moreover, in some applications such as the spectral clustering problem, a superposition of vectors that are forming a solution space for the problem can be used as an input state. In that case, the measurement of the output in the solution space yields the solution for the problem. This method can be used efficiently (polynomial time complexity in the number of qubits) when the vectors describing the solution space are tensor product of Pauli matrices.
}
\section{Complexity Analysis}
\label{section4}
The computational complexity of a quantum algorithm is assessed by the total number of single gates and two qubit controlled NOT (CNOT) gates in the quantum circuit implementing the algorithm. 
We derive the computational complexity of the whole method by finding the complexities of $U_f$ on the first register with $m$ number qubits and $U_{\psi}$ on the second register with $n$ number of qubits.
 We shall use $M=2^m$ and $N=2^n$ to describe the sizes of the operators on the registers.
 
\subsection{The complexity of $U_f$}
It is known that the number of quantum gates to implement a Householder matrix is bounded by the size of the matrix Ref. 
\cite{Ivanov2006engineering,Urias2015householder,Ivanov2008synthesis,Bullock2005asym}. 
Therefore, the circuit for $U_f$ requires $O(M)$ CNOT gates since it is a Householder transformation formed by the vector \ket{\mathbf f} of size $M$.

\subsection{The complexity of $U_{\psi}$}
$U_{\psi}$ is equal to $U_{PEA}U_0U_{PEA}^\dagger$ in which the total complexity will be typically governed by the complexity of $U_{PEA}$. 
$U_{PEA}$ involves the Fourier transforms, input preparation circuit, and the controlled $U=e^{itW}$ with different $t$ values:
\begin{itemize}
\item The circuits for the quantum Fourier transform and its inverse are well known \cite{Nielsen2010quantum} and can be implemented on the first register in $O(m^2)$.
\item  The input preparation circuit on the second register, $U_{input}$, can be implemented again as a Householder transformation  by using $O(N)$ number of quantum gates. 
(It can be also designed by following Sec. III.B. of Ref.\cite{Daskin2012universal}: 
In that case, for every two vector elements, a controlled rotation gate is used to construct $U_{input}$ with the initial row equal to $\mathbf{x}$; 
thus, $U_{input}\ket{\mathbf{0}}=\ket{\mathbf{x}}$.)
\item The circuit complexity of $U=e^{itW}$ is highly related to the structure of $W$.
When $W$ of order $N$ is sparse enough: i.e., the number of nonzero entries is bounded by some polynomial of the number of qubits, $poly(n)$; 
then $W$ can be simulated by using only $O(poly(n))$ number of quantum gates \cite{Aharonov2003,Berry2007sparse,Andrew2011}. However, when $W$ is not sparse but equal to $\sum \mathbf{x_i}\mathbf{x_i}^T$, then as shown in Sec. \ref{SecCircuitW}, we use Trotter-Suzuki formula 
which yields a product of $(4\kappa)$ number of $U_{\mathbf{x_j}}$ matrices with $1 \leq j\leq \kappa$. 
Since $U_{\mathbf{x_j}}$ can be implemented as a Householder transformation by using $O(N)$ quantum gates, 
$U$ requires $O(\kappa N)$ quantum gates.
\end{itemize}
If we combine all the above terms, the total complexity can be concluded as:
\begin{equation}
O(\kappa N+M).
\end{equation}
 This is linear in  system-size, however, exponential in the number of qubits involved in either one of the registers.
In comparison, any classical method applied to obtain $QQ^T\mathbf{x}$ at least requires $O(N^2)$ time complexity because of the matrix vector multiplication.
Therefore, the quantum model presented here may provide a quadratic speed-up over the classical methods for some applications.
 \textbf{When the weight matrix is sparse or the data is given as a quantum states, it can be implemented in $O(poly(n))$. Therefore, the whole complexity becomes linear in the number of qubits, which may provide an exponential speed-up over the classical algorithms. However, when the weight matrix is not sparse,  the complexity becomes exponential in the number of qubits. The current experimental research by big companies such as Google and IBM aims to build 50 qubit operational quantum computers \cite{Castelvecchi2017quantum}. Because of the limitations of the current quantum computer technology, when the required number of qubits goes beyond $50$, the applications of the algorithm becomes infeasible.}
\section{An Illustrative Example}
\label{SecExample}
Here, we give a simple example to show how the algorithm works:
Let us assume, we have given weights represented by the columns of the following matrix \cite{Abdi1999neuralExample}:
\begin{equation}
X=\frac{1}{10}\times\left(\begin{matrix}
-1& +1\\
-1& -1\\
+1& -1\\
-1& +1\\
\end{matrix}\right),
\end{equation}
where we scale the vectors by $\frac{1}{10}$ so as to make sure that the eigenvalues of $W$ are less than one.
To validate the simulation results, first, $W_{[\infty]}$ is classically computed by following  the singular value decomposition of $X$:
\begin{equation}Q\Phi P^T=  \left(\begin{matrix}
+.5774& 0\\
0& 1\\
-.5774& 0\\
+.5774& 0\\
\end{matrix}\right)\left(\begin{matrix}
.2495& 0\\
0& .14142\\
\end{matrix}\right)\left(\begin{matrix}
-.7071& +.7071\\
-.7071& -.7071\\
\end{matrix}\right).
\end{equation}
Therefore, \begin{equation}
W_{[\infty]}=
QQ^T=\left(\begin{matrix}
+.333& 0&-.333&+.333\\
0& 1&0&0\\
-.333& 0&+.333&-.333\\
+.333& 0&-.333&+.333\\
\end{matrix}\right)
\end{equation}
We use the following Trotter-Suzuki formula \cite{Trotter1959product,Suzuki1976generalized,Hatano2005finding,Poulin2015}  to compute the exponential of $W=XX'$:
\begin{equation}
U=e^{i2\pi W}\approx 
e^{i\pi \mathbf{x_2} \mathbf{x_2^T}}
e^{i2\pi \mathbf{x_1} \mathbf{x_1^T}} e^{i\pi \mathbf{x_2} \mathbf{x_2^T}}
\end{equation}

In the simulation for a random input \ket{\mathbf x}, the comparison of $W_{[\infty]}\ket{\mathbf x}=
QQ^T\ket{\mathbf x}$ with the output of the second register in the quantum model yields the fidelity. 
For two different random inputs, the simulation results in each iteration are shown in Fig.\ref{FigDiffIna} and Fig.\ref{FigDiffInb} for 
$\ket{\mathbf x}=
\left(.3517\ 
    .3058\ 
    .6136\ 
    .6374\right)^T$ and  
    $\ket{\mathbf x}=
(.7730\
    .1919\
    .1404\
    .5881)^T$, respectively.

\section{Conclusion}
The weight matrix of the networks based on the Widrow-Hoff learning rule converges to $QQ^T$, where $Q$ represents the eigenvectors of the matrix corresponding to the nonzero eigenvalues. 
In this paper, we have showed how to apply the quantum principal component analysis method described in Ref.\cite{Daskin2016QPCA} to artificial neural networks using the Widrow-Hoff learning rule. 
In particular, 
we have show that one can implement an equivalent quantum circuit which produces the output $QQ^T\mathbf{x}$ for a given input $\mathbf{x}$ in linear time.
The implementation details are discussed by using random cases, and the computation complexity is analyzed based on the number of quantum gates. 
In addition, a simple numerical example is presented. 
The model is general and requires only linear time computational complexity in the size of the weight matrix.

\appendix{
\section{ MATLAB Code for the Random Matrix}
\label{AppRandom}
The random matrix used in the numerical example is generated by the following MATLAB code snippet:
\begin{verbatim}
%number of non-zero eigenvalues
npc = ceil(N/2);
d = rand(N,1);%random eigenvalues
d(npc+1:end) = 0;
%random eigenvectors
[Qfull,~] = qr(randn(N));
%the unitary matrix in PEA
U = Qfull*diag(exp(1i*2*pi*d))*Qfull';
%normalized input vector
x = rand(N,1); x = x/norm(x);
\end{verbatim}
}


\pagebreak
\begin{figure*}
\subfloat[m is 1.]{
\includegraphics[width=3.5in]{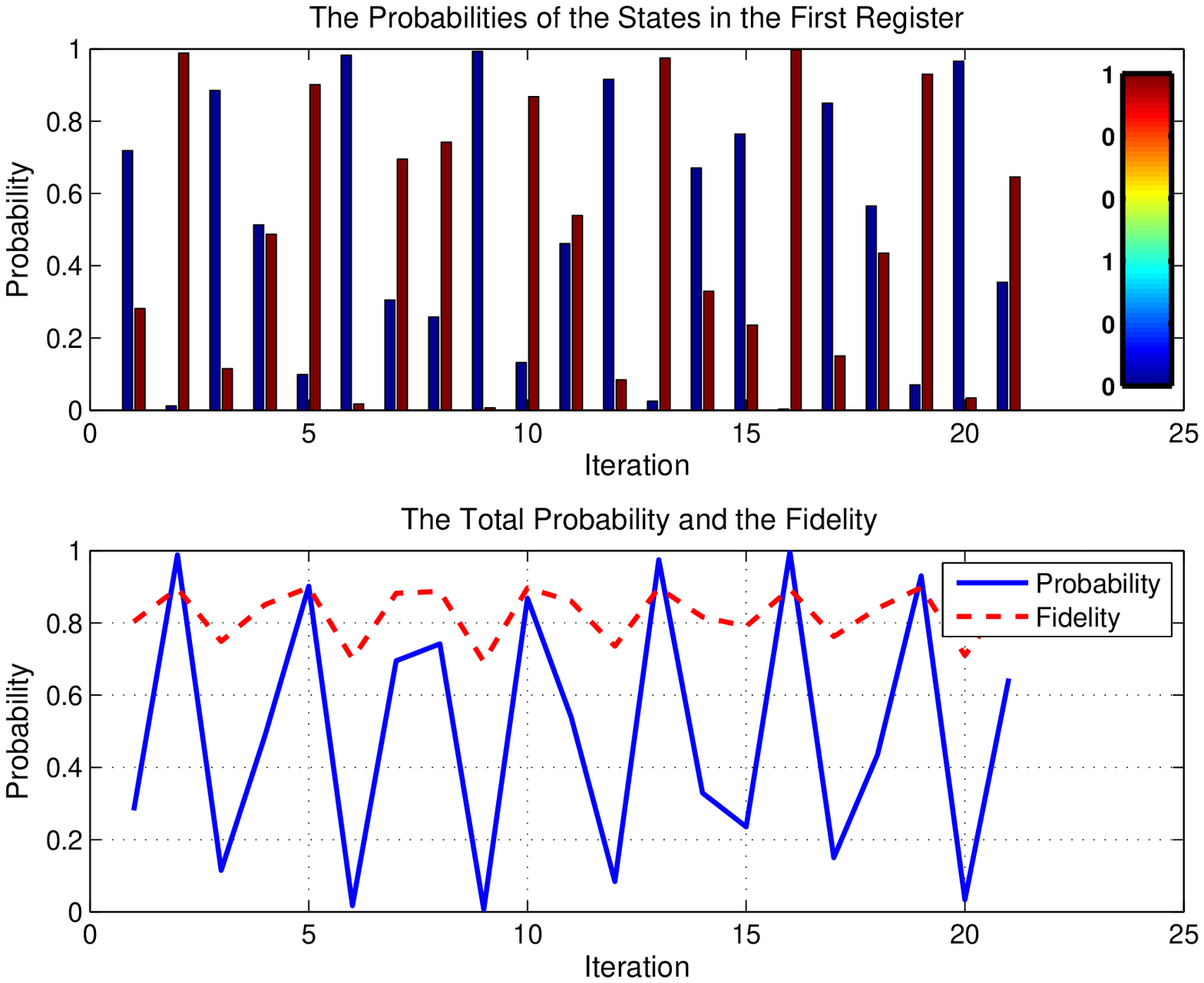}
}
\subfloat[$m$ is 2.]{
\includegraphics[width=3.5in]{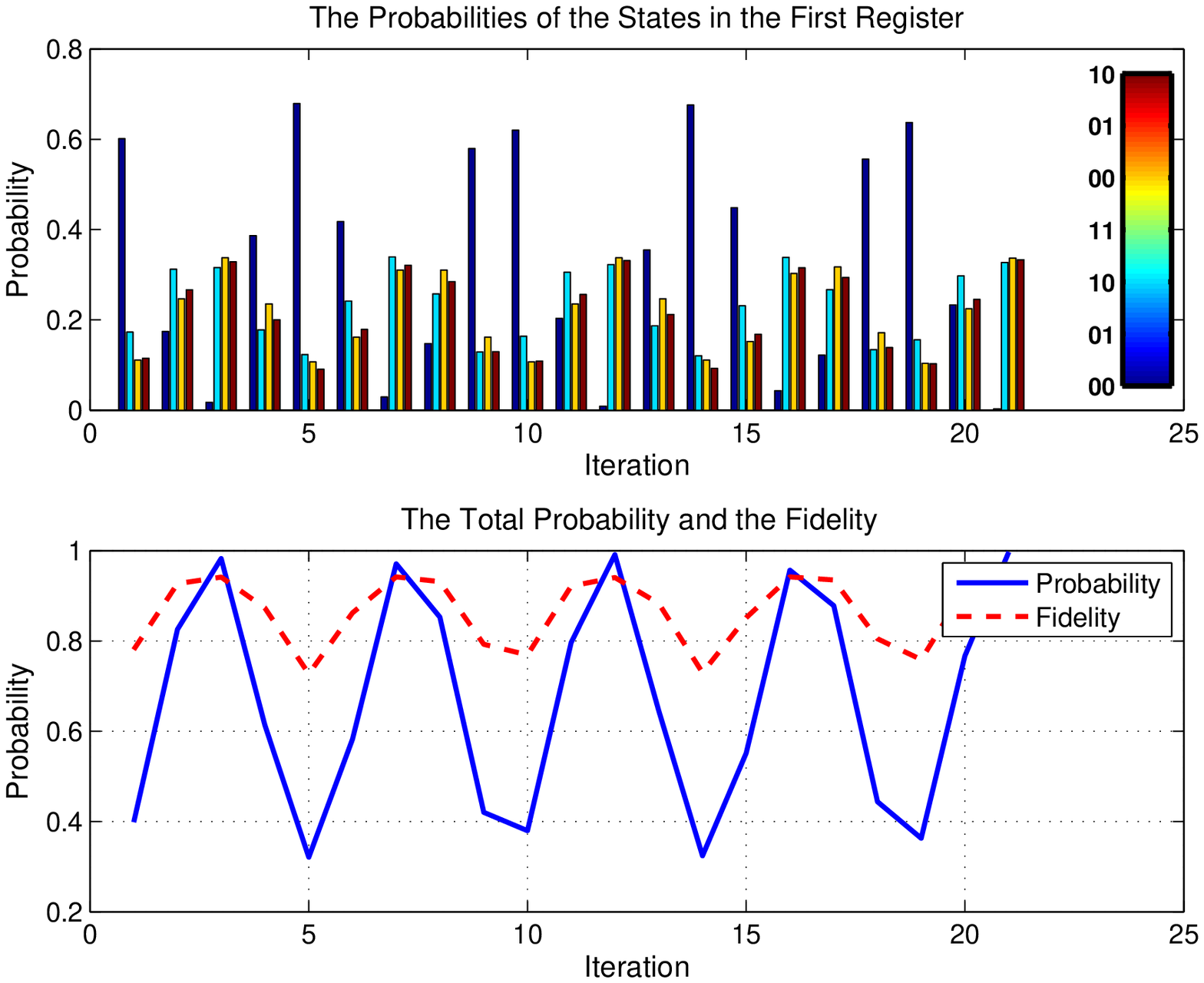}
}\\
\subfloat[$m$ is 3.]{
\includegraphics[width=3.5in]{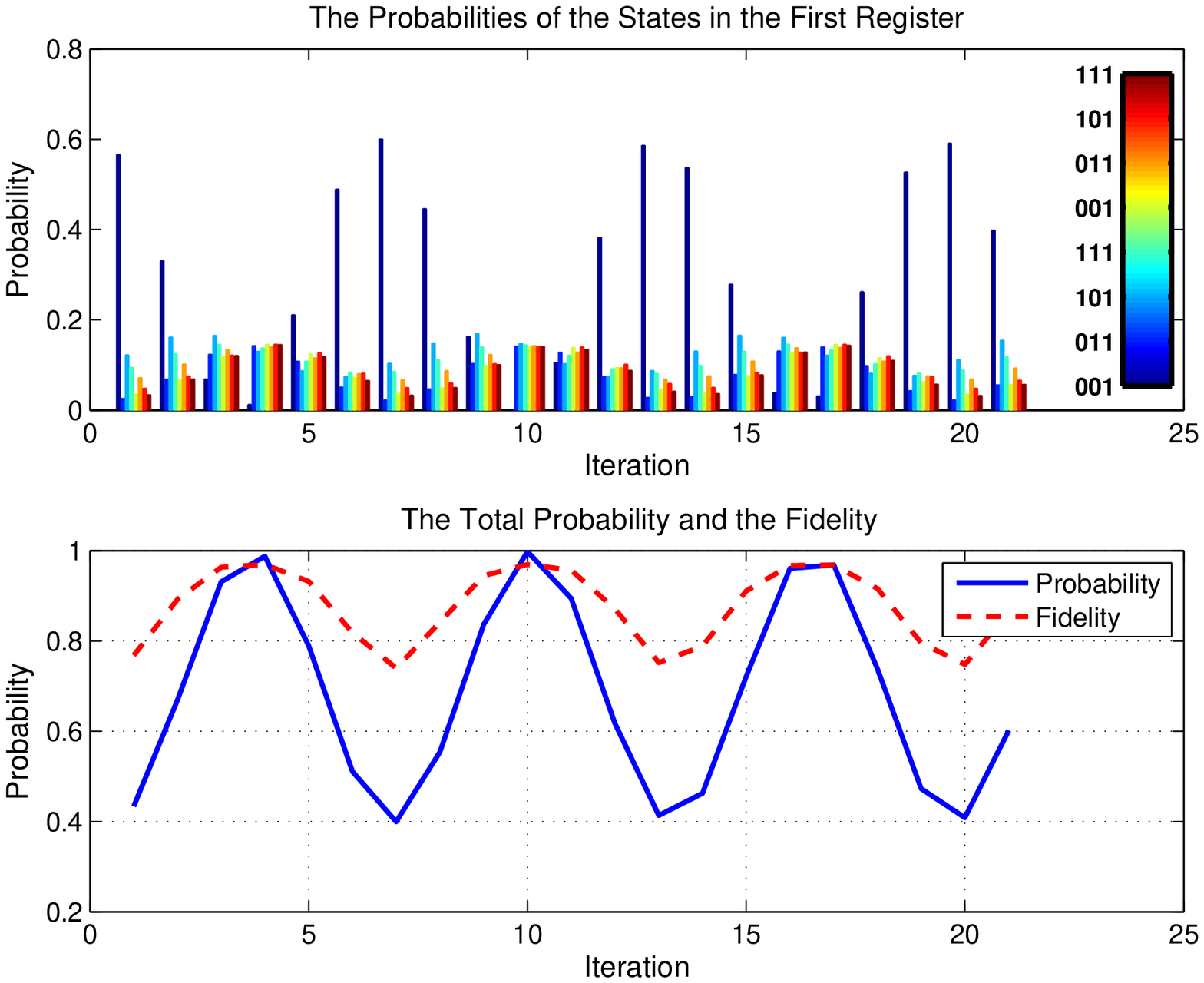}
}
\subfloat[$m$ is 4.]{
\includegraphics[width=3.5in]{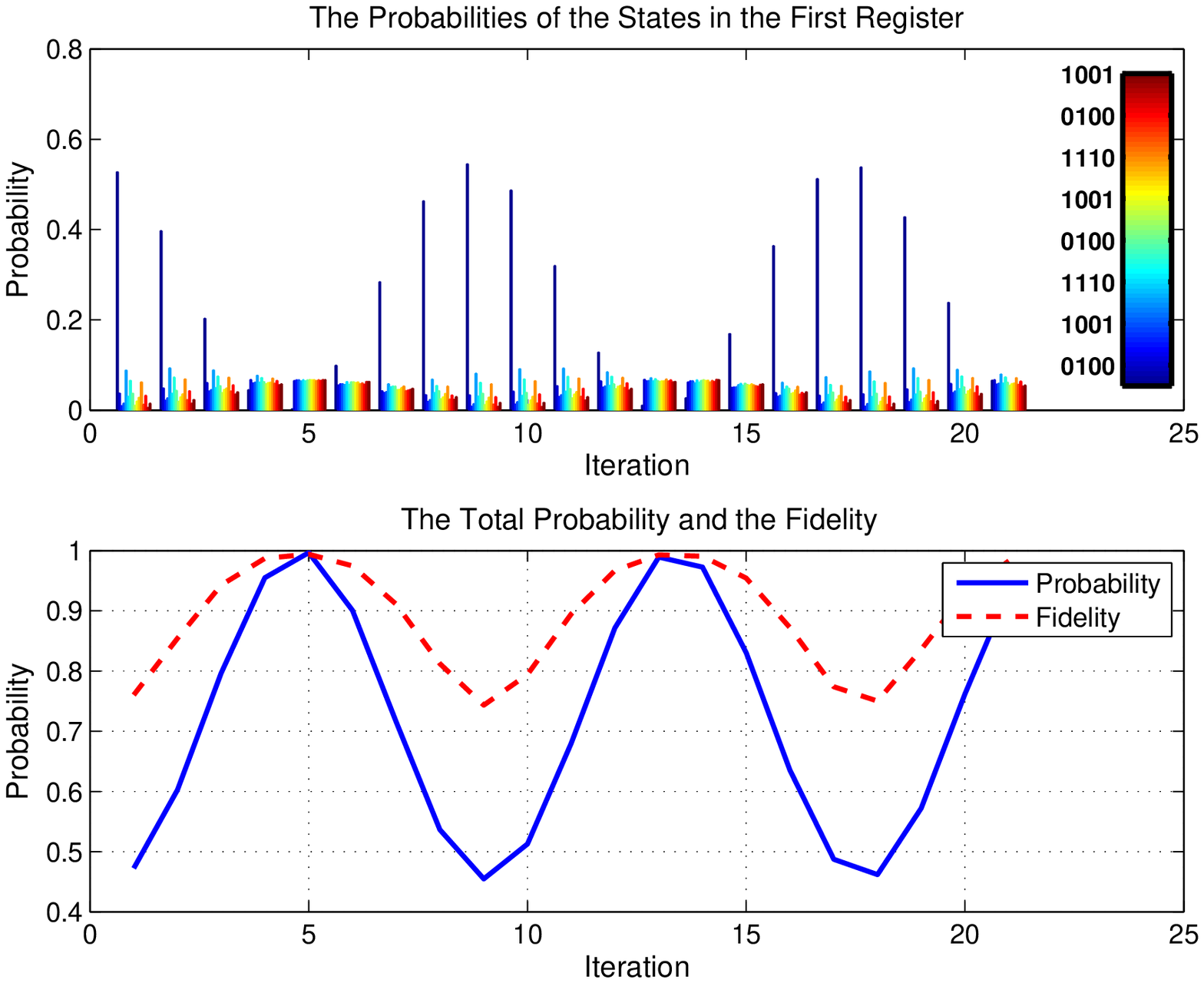}
}\\
\subfloat[$m$ is 5.]{
\includegraphics[width=3.5in]{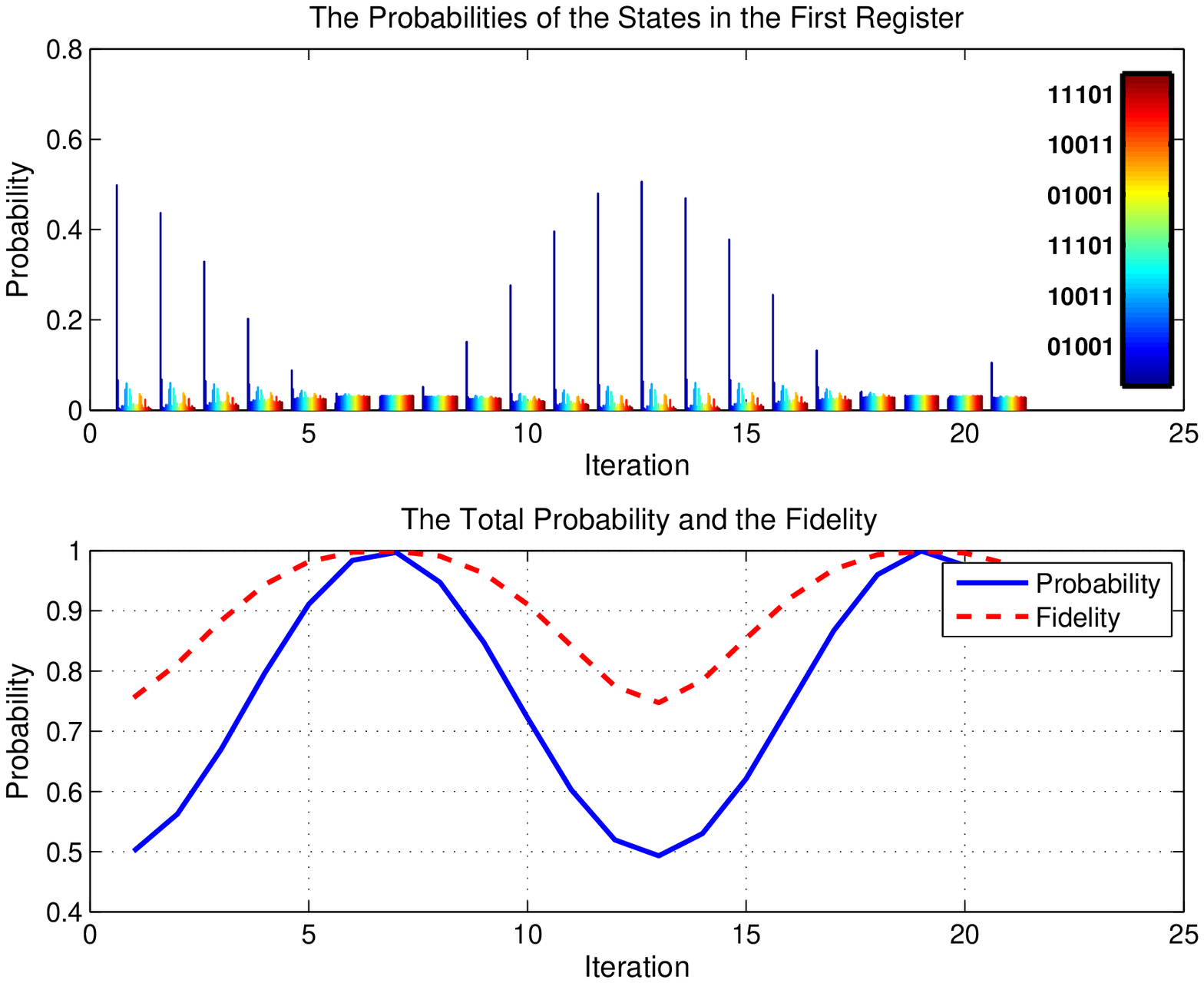}
}
\subfloat[$m$ is 6.]{\label{FigRandomQANNm6}
\includegraphics[width=3.5in]{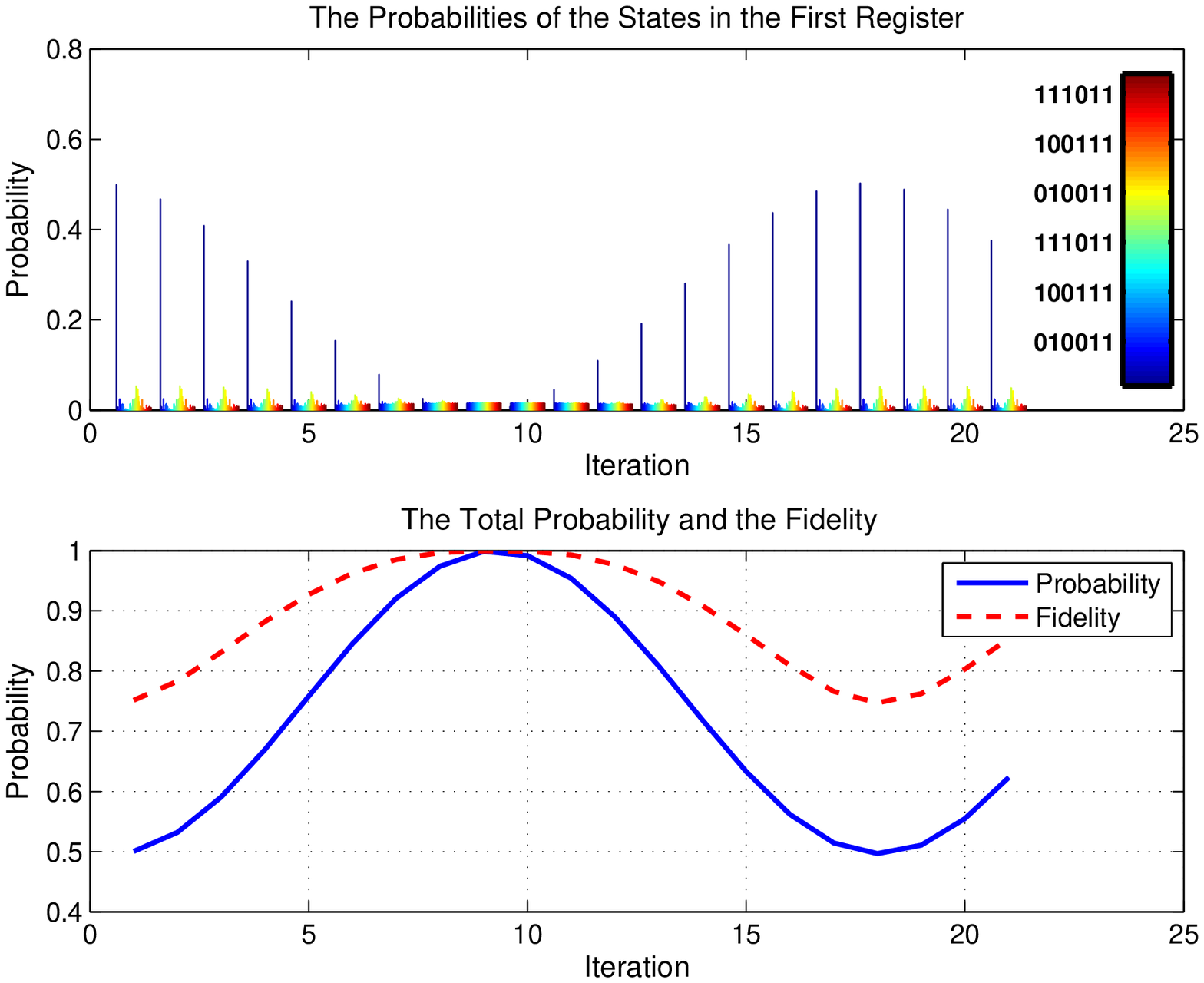}
}
\caption{The probability changes in the iteration of the amplitude amplification for a random $2^7 \times 2^7$  matrix with $2^7/2$ number of zero eigenvalues and a random input $\ket{\mathbf x}$ 
(MATLAB code for the random generation is given in Appendix \ref{AppRandom}). 
In each subfigure, we have used different numbers of qubits, $m$, for the first register to see the effect on the results. 
The bar graphs in the subfigures shows the probability change for each state \ket{\mathbf j}, $j=0\dots 1$, of the first qubit. For each state, a different color tone is used.}
\label{FigRandomQANN}
\end{figure*}

\begin{figure*}
\subfloat[\label{FigDiffIna}For the generated random input 
$\ket{\mathbf x}=(
.3517\  
    .3058\  
    .6136\  
    .6374)^T$.]{
\includegraphics[width=3.5in]{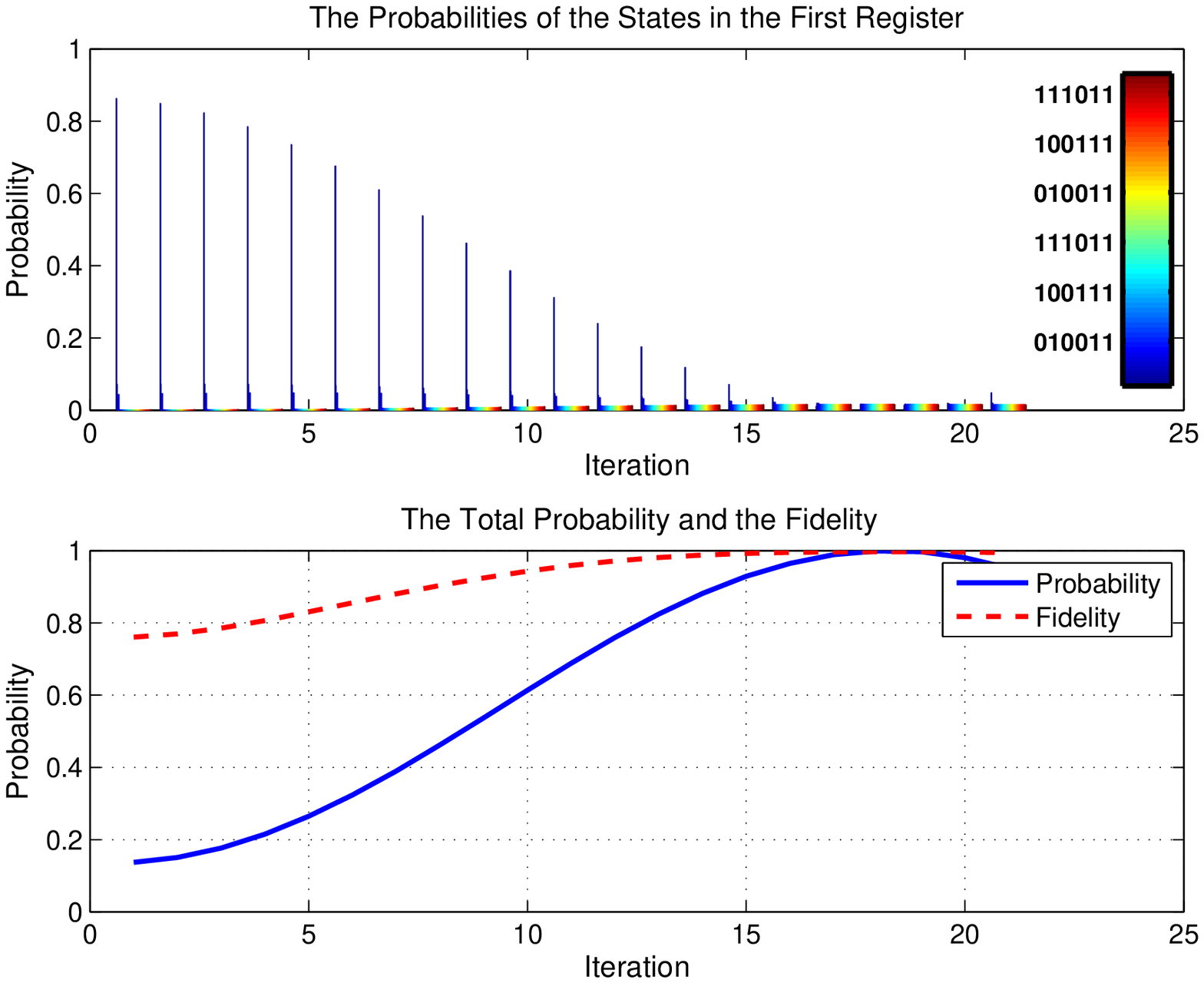}
}
\subfloat[For the generated random input 
$\ket{\mathbf x}=
(.7730\
    .1919\
    .1404\
    .5881)^T$.]{
\includegraphics[width=3.5in]{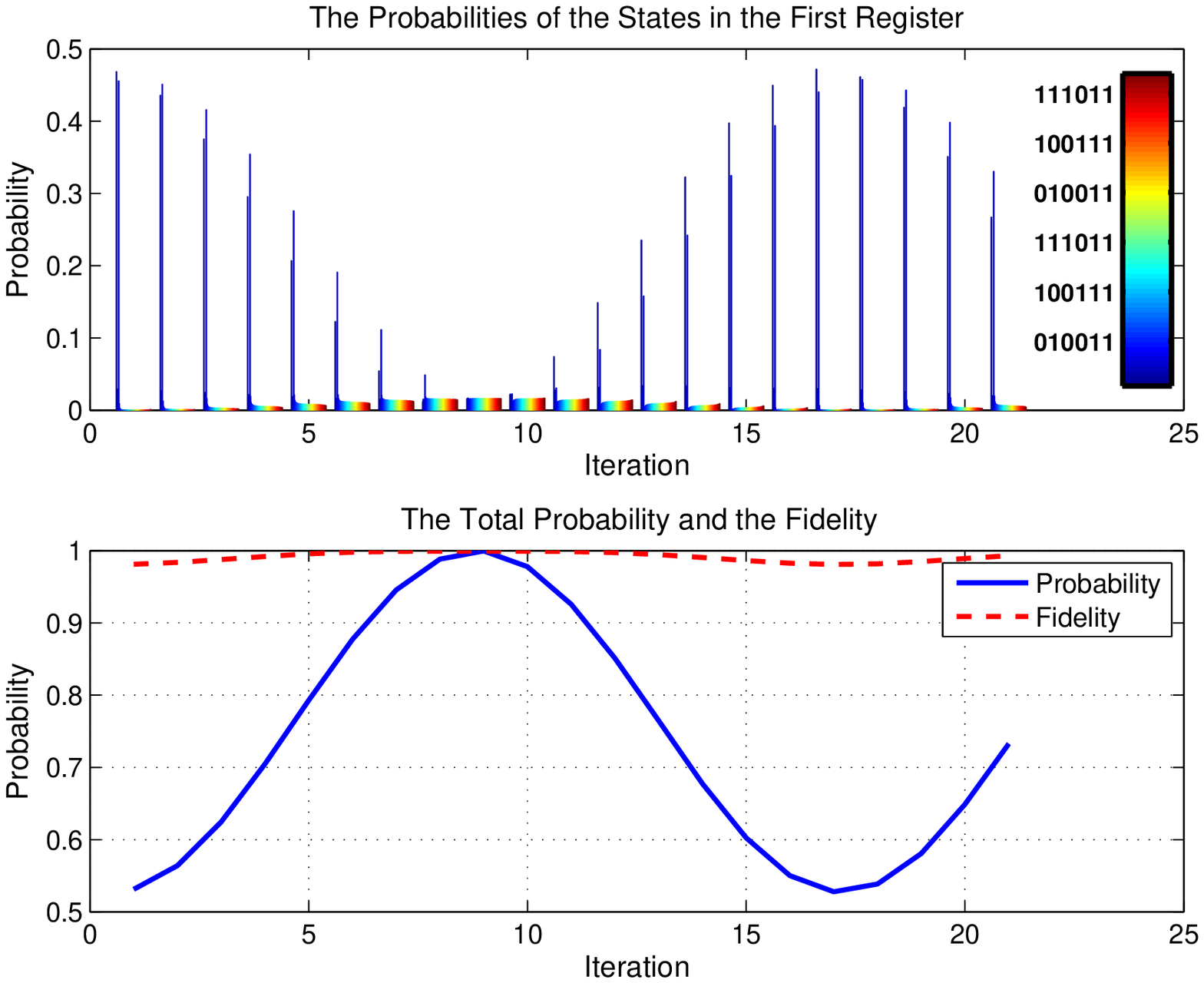}\label{FigDiffInb}
}
\caption{The simulation results of the quantum model for the example in Sec.\ref{SecExample} with two different input vectors. }
\label{FigDiffIn}
\end{figure*}
\end{document}